\documentclass[showpacs,pra]{revtex4}
\usepackage[latin9]{inputenc}
\usepackage{graphicx}
\usepackage{esint}

\makeatletter
\@ifundefined{textcolor}{}
{%
 \definecolor{BLACK}{gray}{0}
 \definecolor{WHITE}{gray}{1}
 \definecolor{RED}{rgb}{1,0,0}
 \definecolor{GREEN}{rgb}{0,1,0}
 \definecolor{BLUE}{rgb}{0,0,1}
 \definecolor{CYAN}{cmyk}{1,0,0,0}
 \definecolor{MAGENTA}{cmyk}{0,1,0,0}
 \definecolor{YELLOW}{cmyk}{0,0,1,0}
}

\usepackage{epsfig}\usepackage{color}

\makeatother

\begin{document}

\title{Generation of helical modes from a topological defect}

\author{ Sébastien Fumeron$^{1}$, Fernando Moraes$^{1,2}$ and Erms Pereira$^{3}$.}

\affiliation{$^{1}$ Laboratoire d'Énergétique et de Mécanique Théorique et Appliquée\\
 CNRS UMR 7563\\
 Nancy Université\\
 54506 Vandoeuvre Cedex, France.}

\affiliation{$^{2}$ Departamento de Física, CCEN, Universidade Federal da Paraíba,
\\
 Caixa Postal 5008, 58051-970 João Pessoa, PB, Brazil}

\affiliation{$^{3}$ Instituto de Física, Universidade Federal de Alagoas, \\
57072-900, Maceió, Alagoas, Brazil.}
\begin{abstract}
{The propagation of an electromagnetic wave in a medium with a screw
dislocation is studied. Adopting the formalism of differential forms,
it is shown that torsion is responsible for quantized modes. Moreover,
it is demonstrated that the modes thus obtained have well defined
orbital angular momentum, opening the possibility to design liquid-crystal-based
optical tweezers.} 
\end{abstract}


\maketitle

\section{Introduction}

During the last decades, the interaction of the orbital angular momentum
of light with matter has become a very active research field \cite{muthukrishnan,babiker,alexandrescu,thanvanthri}
due to its large number of potential applications such as optical
tweezers (for the manipulation of living cells and nanoobjects), micromachines
(molecular engines) or quantum cryptography devices. In electrodynamics,
it is indeed well known that light carries an angular momentum. This
latter can be divided into two parts: a spin contribution associated
to the polarization of the wave and an orbital contribution. A possible
way of controling the orbital angular momentum state of light beams
is to use q-plates built out of liquid crystals \cite{marrucci}.
These q-plates coincide with the cross sections of topological line
defects \cite{satiro} and therefore, understanding how the angular
momentum of light interacts with a line defect is of prime interest.

From this point of view, the most relevant line defects are probably
screw dislocations, because they locally induce torsion. A screw dislocation
is a line defect that may occur in smectic C{*} liquid crystals \cite{achard},
in ordinary crystals \cite{nabarro} and even in spacetime \cite{galtsov}.
The generation of the defected topology is achieved through a ``cut
and glue'' Volterra process, based on ideas of the homology theory
\cite{kleman}: basically, the screw dislocation is generated by cutting
the medium along a half-plane, moving the part located over the cut
by a vector $\vec{b}$ (named Burgers vector) parallelwise to the
edge of the cutting plane, and finally gluing the upper and lower
sides. Thus, a screw dislocation is associated with a breaking of
translational symmetry and it also exhibits an explicit helicity which,
as we show below, has a profound influence on the angular momentum
of a propagating electromagnetic field. Fig. 1 depicts a screw dislocation
in a generic continuous medium. Assuming cylindrical coordinates and
taking the axis of the defect to be the $z$-axis, it is clear that
the screw dislocation mixes the $r$ and $\varphi$ degrees of freedom.
In other words, by going clockwise a complete turn around the axis,
one moves up by one unit of Burgers vector $\vec{b}$.

\begin{figure}[!h]
\centering{}\includegraphics[height=5cm]{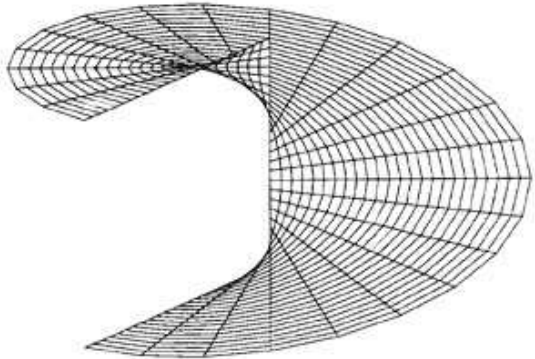} \caption{Screw dislocation}
\end{figure}

\noindent An elegant way of taking into account the boundary condition
\begin{eqnarray}
\varphi\rightarrow\varphi+2\pi & implies & z\rightarrow z+b\label{condition}
\end{eqnarray}
is to use an Einstein-Cartan background \cite{puntigam}. This approach
has also been used to describe elastic continuous media in analogy
with gravity \cite{katanaev}. In this work, we look for a simple
solution for an electromagnetic wave propagating along the axis of
a screw dislocation. Since the main purpose of this article is to
demonstrate the acquisition of orbital angular momentum by the propagating
fields, this work may be relevant for applications both in condensed
matter physics (particularly smectic C{*} liquid crystals) and also
in cosmology. For example, searching the cosmic microwave background
for orbital angular momentum beams could give some clues on the existence
of cosmic screw dislocations in the early universe. For simplicity
we consider $c=\varepsilon=\mu=1$. Even though the problem at hand
involves non-relativistic systems, for convenience, we work in a four-dimensional
spacetime as it provides a framework in which Maxwell's equations
are naturally covariant. In cylindrical coordinates, the background
geometry induced by the screw dislocation is given by the line element
\cite{galtsov,puntigam} 
\begin{equation}
ds^{2}=-dt^{2}+dr^{2}+r^{2}d\varphi^{2}+(dz+\beta d\varphi)^{2},\label{metric}
\end{equation}
where $\beta=b/2\pi$. Explicitly, the metric tensor is therefore
\begin{equation}
g_{\mu\nu}=\left(\begin{array}{cccc}
-1 & 0 & 0 & 0\\
0 & 1 & 0 & 0\\
0 & 0 & r^{2}+\beta^{2} & \beta\\
0 & 0 & \beta & 1
\end{array}\right).\label{g-quiral}
\end{equation}

It must be emphasized that for solid crystals, this geometrization
of matter is actually qualitatively equivalent to determinations of
actual properties of screw-dislocated dielectrics based on usual elasticity
theory. An elastic defect is indeed expected to modify the dielectric
properties in the vicinity of the dislocation as prescribed in eq.
(11) of reference \cite{sahoo}. This latter has to be compared to
the spatial part of the metric (\ref{g-quiral}) rewritten in Cartesian
coordinates, that is 
\begin{equation}
g_{ij}=\delta_{ij}+\frac{b}{2\pi r^{2}}\left(\begin{array}{ccc}
0 & 0 & -y\\
0 & 0 & x\\
-y & x & 0
\end{array}\right).\label{cartesian}
\end{equation}
up to first order in $\frac{b}{2\pi r^{2}}$. Therefore, it is clear
that the anisotropy introduced in the dielectric medium by the screw
dislocation, to a good approximation, can effectively be described
by a background space with unit dielectric constant given by metric
(\ref{cartesian}). However, this is done in a qualitative way since
the coupling constant $P_{2}$, between the strain field and the electromagnetic
field, does not appear explicitly in our model. This is due to the
fact that the starting point of the geometric approach is the boundary
condition (\ref{condition}) and not the elasto-optic effect.

On the other hand, in order to describe electromagnetic waves propagating
along the axis of a cosmic screw dislocation, we assume that there
is no other source of geometry (gravitational field) in the vicinity
of the defect. Also, the dislocation is supposed not to be rotating,
which would include a coupling between $t$ and $\varphi$ in the
metric, just like the one between $z$ and $\varphi$ due to the dislocation.
Moreover, we consider a fixed background geometry, that is, we assume
that the electromagnetic wave energy contribution to the local gravitational
field is negligible. In either case of propagating electromagnetic
fields along a screw dislocation, be it in condensed matter or in
the cosmos, we have a geometrical background given by (\ref{metric}).
Once the metric tensor is known, the language of General Relativity
provides a powerful tool to determine the equations governing electrodynamics
in the distorted background. This is the object of the next section.

\section{Maxwell's equations}

In reference \cite{dias}, Maxwell's equations were found in the geometry
induced by the presence of a cosmic dislocation using the differential
forms formalism \cite{warnick,baez}. Besides its natural elegance,
the main advantage of this formalism is the fact that it provides
a coordinate-free formulation of electrodynamics. Coordinates are
introduced only when a specification of the field components is required.
In what follows, the derivation of Maxwell's equations in the screw-dislocated
background is presented, following the steps of reference \cite{dias}.
From the point of view of spacetime, we consider the approximation
\cite{petterson} where the electromagnetic field is taken as a weak
perturbation on the spacetime metric. That way, the contribution of
the electromagnetic field to the spacetime geometry is neglected and
the Einstein-Maxwell equations are decoupled. From the point of view
of solid-state physics, this approximation means that the electromagnetic
field is supposed not to affect the elastic properties of the material
medium.

In language of differential forms, Maxwell's equations can be concisely
expressed as \cite{baez}

\begin{equation}
\mbox{\boldmath\ensuremath{d}}\mbox{\boldmath\ensuremath{F}}=0\label{dF=00003D0}
\end{equation}
and

\begin{equation}
\star\mbox{\boldmath\ensuremath{d}}\!\star\!\mbox{\boldmath\ensuremath{F}}=\mbox{\boldmath\ensuremath{J}}.\label{*d*F=00003DJ}
\end{equation}
Here, \textbf{\textsl{d}} denotes the exterior derivative, $\star$
is the Hodge star operator (see Appendix), \textbf{F} is the Faraday
2-form defined as: 
\begin{equation}
\mbox{\boldmath\ensuremath{F}}\equiv\mbox{\boldmath\ensuremath{B}}+\mbox{\boldmath\ensuremath{E}}\wedge\mbox{\boldmath\ensuremath{d}}t.\label{defF}
\end{equation}

\noindent and $\mbox{\boldmath\ensuremath{J}}$ is the current density
1-form 
\begin{equation}
\mbox{\boldmath\ensuremath{J}}=-\rho\mbox{\boldmath\ensuremath{d}}t+J_{r}\mbox{\boldmath\ensuremath{d}}r+J_{\varphi}\mbox{\boldmath\ensuremath{d}}\varphi+J_{z}\mbox{\boldmath\ensuremath{d}}z,
\end{equation}

\noindent The electric field 1-form is written as 
\begin{equation}
\mbox{\boldmath\ensuremath{E}}=E_{r}\mbox{\boldmath\ensuremath{d}}r+E_{\varphi}\mbox{\boldmath\ensuremath{d}}\varphi+E_{z}\mbox{\boldmath\ensuremath{d}}z,
\end{equation}

\noindent and the magnetic field 2-form is given by

\begin{equation}
\mbox{\boldmath\ensuremath{B}}=B_{\phi z}\mbox{\boldmath\ensuremath{d}}\phi\wedge\mbox{\boldmath\ensuremath{d}}z+B_{zr}\mbox{\boldmath\ensuremath{d}}z\wedge\mbox{\boldmath\ensuremath{d}}r+B_{r\phi}\mbox{\boldmath\ensuremath{d}}r\wedge\mbox{\boldmath\ensuremath{d}}\phi.\label{2formaB-cil}
\end{equation}

In order to express Maxwell's equations in terms of the electric and
magnetic field components of the usual Euclidean space, it is necessary
to find the transformation laws between the components of a differential
form and its components in the Euclidean basis. The vector basis of
3-dimensional Euclidean space is the space subset of $\mathcal{B}_{\hat{v}}=\{\vec{e}_{\hat{t}},\vec{e}_{\hat{r}},\vec{e}_{\hat{\varphi}},\vec{e}_{\hat{z}},\}$
such that $\vec{e}_{\hat{\mu}}\cdot\vec{e}_{\hat{\nu}}=\eta_{\mu\nu}$,
where $\eta_{\mu\nu}$ is the flat Minkowski metric. The above basis
is not $\mathcal{B}_{v}=\{\vec{e}_{t},\vec{e}_{r},\vec{e}_{\varphi},\vec{e}_{z}\}$,
the dual basis of $\mathcal{B}_{1}$, which is such that $\vec{e}_{\mu}\cdot\vec{e}_{\nu}=g_{\mu\nu}$.
The relation between the vectors in the basis $\mathcal{B}_{v}$ and
$\mathcal{B}_{\hat{v}}$ is therefore:

\begin{equation}
\vec{e}_{\hat{\mu}}=\frac{\vec{e}_{\mu}}{\sqrt{|g_{\mu\mu}|}}.\label{rel-bases}
\end{equation}

\noindent As a consequence, the transformation law between the Euclidean
components of a vector $\vec{v}$ and its contravariant components
are related by \cite{foot}

\begin{equation}
v^{\mu}=\frac{v^{\hat{\mu}}}{\sqrt{|g_{\mu\mu}|}}.\label{rel-componentes}
\end{equation}
(Notice that, in the last two equations, the sum convention for repeated
indices should not be used.) Using the metric to obtain the dual vectors
of the 1-form (that is the contravariant vectors) and (\ref{rel-componentes}),
one finally gets the transformation laws between the components of
a 1-form $\mbox{\boldmath\ensuremath{A}}$ and its components in the
Euclidean basis: 
\begin{equation}
A_{r}=A^{\hat{r}},~~A_{\varphi}=\sqrt{\alpha^{2}r^{2}+\beta^{2}}~A^{\hat{\varphi}}+\beta A^{\hat{z}}~~\mbox{e}~~A_{z}=A^{\hat{z}}+\frac{\beta}{\sqrt{\alpha^{2}r^{2}+\beta^{2}}}A^{\hat{\varphi}}.
\end{equation}

After some algebra, one finally obtains Maxwell's equation in the
presence of the screw dislocation as \cite{dias}:

\begin{equation}
\frac{1}{r}\frac{\partial}{\partial r}(rE^{\hat{r}})+\frac{1}{\sqrt{r^{2}+\beta^{2}}}\frac{\partial E^{\hat{\varphi}}}{\partial\varphi}+\frac{\partial E^{\hat{z}}}{\partial z}=\rho,\label{gauss-quiral}
\end{equation}
for Gauss' law, whereas for the three components of Ampère-Maxwell's
law, it comes: 
\begin{equation}
\frac{1}{r}\left[\left(\frac{\beta}{\sqrt{r^{2}+\beta^{2}}}\frac{\partial}{\partial\varphi}-\sqrt{r^{2}+\beta^{2}}\frac{\partial}{\partial z}\right)B^{\hat{\varphi}}+\left(\frac{\partial}{\partial\varphi}-\beta\frac{\partial}{\partial z}\right)B^{\hat{z}}\right]=J^{\hat{r}}+\frac{\partial E^{\hat{r}}}{\partial t},\label{r-ampere-quiral}
\end{equation}

\begin{equation}
\frac{\sqrt{r^{2}+\beta^{2}}}{r}\left[\frac{\partial B^{\hat{r}}}{\partial z}-\frac{\partial}{\partial r}\left(B^{\hat{z}}+\frac{\beta}{\sqrt{r^{2}+\beta^{2}}}B^{\hat{\varphi}}\right)\right]=J^{\hat{\varphi}}+\frac{\partial E^{\hat{\varphi}}}{\partial t},\label{fi-ampere-quiral}
\end{equation}

\begin{equation}
\frac{1}{r}\left[\frac{\partial}{\partial r}\left(\sqrt{r^{2}+\beta^{2}}~B^{\hat{\varphi}}+\beta B^{\hat{z}}\right)-\frac{\partial B^{\hat{r}}}{\partial\varphi}\right]=J^{\hat{z}}+\frac{\partial E^{\hat{z}}}{\partial t}.\label{z-ampere-quiral}
\end{equation}

\noindent The equations describing the absence of magnetic monopoles
and Faraday's law are both obtained by making 
\begin{eqnarray*}
\rho=0,~~J^{\hat{r}}=J^{\hat{\varphi}}=J^{\hat{z}}=0,\\
B^{\hat{i}}\rightarrow E^{\hat{i}}~~~~e~~~~E^{\hat{i}}\rightarrow-B^{\hat{i}},
\end{eqnarray*}
where $i$ corresponds to the indices $r$, $\varphi$ and $z$. The
absence of magnetic monopoles leads to 
\begin{equation}
\frac{1}{r}\frac{\partial}{\partial r}(rB^{\hat{r}})+\frac{1}{\sqrt{r^{2}+\beta^{2}}}\frac{\partial B^{\hat{\varphi}}}{\partial\varphi}+\frac{\partial B^{\hat{z}}}{\partial z}=0,\label{monopolo-quiral}
\end{equation}
and the three components of Faraday's law are given by: 
\begin{equation}
\frac{1}{r}\left(\frac{\beta}{\sqrt{r^{2}+\beta^{2}}}\frac{\partial}{\partial\varphi}-\sqrt{r^{2}+\beta^{2}}\frac{\partial}{\partial z}\right)E^{\hat{\varphi}}+\frac{1}{r}\left(\frac{\partial}{\partial\varphi}-\beta\frac{\partial}{\partial z}\right)E^{\hat{z}}=-\frac{\partial B^{\hat{r}}}{\partial t},\label{r-faraday-quiral}
\end{equation}

\begin{equation}
\frac{\sqrt{r^{2}+\beta^{2}}}{r}\left[\frac{\partial E^{\hat{r}}}{\partial z}-\frac{\partial}{\partial r}\left(E^{\hat{z}}+\frac{\beta}{\sqrt{r^{2}+\beta^{2}}}E^{\hat{\varphi}}\right)\right]=-\frac{\partial B^{\hat{\varphi}}}{\partial t},\label{fi-faraday-quiral}
\end{equation}

\begin{equation}
\frac{1}{r}\left[\frac{\partial}{\partial r}\left(\sqrt{r^{2}+\beta^{2}}~E^{\hat{\varphi}}+\beta E^{\hat{z}}\right)-\frac{\partial E^{\hat{r}}}{\partial\varphi}\right]=-\frac{\partial B^{\hat{z}}}{\partial t}.\label{z-faraday-quiral}
\end{equation}

\section{An heuristic solution}

Except for the plane wave, most of the propagating solutions of Maxwell's
equations are of quite complicated form. They are usually described
as superpositions of plane waves or, depending on the coordinate system,
special functions or polynomial expansions. Since we are interested
in orbital angular momentum, it would appear natural to look for solutions
of the Laguerre-Gaussian beam type \cite{allen}, for example. Nevertheless,
the aim of this article, being pedagogical, is to find a simple propagating
solution that contains the essential physics of the system.

We consider a wave propagating along the Burgers vector of the defect.
Therefore, it is licit to assume that variables can be separated according:
\begin{eqnarray}
\vec{E} & = & \vec{E}_{0}(r)u(\varphi)\exp\left[ikz-i\omega t\right],\nonumber \\
\vec{B} & = & \vec{B}_{0}(r)u(\varphi)\exp\left[ikz-i\omega t\right],\label{general_ansatz}
\end{eqnarray}
whith $u(\varphi)$ a complex-valued function. That way, after some
calculations, (\ref{gauss-quiral}) writes as the sum of two terms:
\begin{equation}
\frac{\sqrt{r^{2}+\beta^{2}}}{E_{0}^{\hat{\varphi}}}\left[\frac{1}{r}\frac{\partial}{\partial r}(rE_{0}^{\hat{r}})+ikE_{0}^{\hat{z}}\right]+\frac{1}{u(\varphi)}\frac{du}{d\varphi}=0,\label{separ_var}
\end{equation}
The first term depends only on \textit{r} whereas the second depends
only on $\varphi$. As a consequence, it is mandatory that: 
\begin{equation}
\frac{1}{u(\varphi)}\frac{du}{d\varphi}=C,\label{separ_var2}
\end{equation}
with $C$ being a constant complex number. Moreover, for symmetry
reasons, it is required that under the transformation $\phi\rightarrow\phi+2\pi,z\rightarrow z+b$,
the field remains unchanged so that finally: 
\begin{equation}
u(\varphi)=\exp\left[id\varphi\right],\label{form_u}
\end{equation}
with \textit{d} a real number such that $\nu=id$. As a consequence,
the field is expected to have the following form: 
\begin{eqnarray}
\vec{E} & = & \vec{E}_{0}(r)\exp\left[ikz+id\varphi-i\omega t\right],\nonumber \\
\vec{B} & = & \vec{B}_{0}(r)\exp\left[ikz+id\varphi-i\omega t\right],\label{ansatz}
\end{eqnarray}
where $k$ is the wavevector along the $z$ direction, $d$ a real
number, and $\omega$ the angular frequency.

By direct substitution of (\ref{ansatz}) into the Maxwell's equations
(\ref{gauss-quiral})-(\ref{z-faraday-quiral}), separating the real
and imaginary parts of each equation, and solving the resulting system
of equations, it comes that: 
\begin{equation}
\vec{E}_{0}(r)=\frac{a}{r}\vec{e}_{\hat{r}}-a\frac{\sqrt{r^{2}+\beta^{2}}}{r^{2}}\vec{e}_{\hat{\varphi}}+\frac{\beta a}{r^{2}}\vec{e}_{\hat{z}}\label{Efield}
\end{equation}
and 
\begin{equation}
\vec{B}_{0}(r)=\frac{a}{r}\vec{e}_{\hat{r}}+a\frac{\sqrt{r^{2}+\beta^{2}}}{r^{2}}\vec{e}_{\hat{\varphi}}-\frac{\beta a}{r^{2}}\vec{e}_{\hat{z}},\label{Bfield}
\end{equation}
where the parameter $a$ sets the intensity of the fields. We also
get the dispersion relation 
\begin{equation}
k=\pm\omega\label{dispersion}
\end{equation}
but more importantly that the integer $m$ is related to the Burgers
vector and to the wavevector by 
\begin{equation}
m=\beta k.\label{eme}
\end{equation}
(Notice that the solution (\ref{ansatz}) satisfies the boundary condition
(\ref{condition}) since equation (\ref{eme}) holds). This implies
that solutions of the type (\ref{ansatz}) are quantized, that is
only modes with definite wavevector $k_{m}=m/\beta$ are allowed.
This brings about interesting applications such as using the medium
endowed with a defect as a filter for specific frequencies.

From equations (\ref{Efield}) and (\ref{Bfield}), we obtain the
Poynting vector 
\begin{equation}
\vec{S}=\frac{1}{2}\vec{E}\times\vec{B}^{*}=\frac{\beta a^{2}}{r^{3}}\vec{e}_{\hat{\varphi}}+\frac{a^{2}\sqrt{r^{2}+\beta^{2}}}{r^{3}}\vec{e}_{\hat{z}}.\label{poynting}
\end{equation}

Moreover, with only $\varphi$ and $z$ components, it appears that
the Poynting vector spirals along the direction of propagation. To
verify this, we identify its components with the components of a tangent
vector to a yet unknown space curve given in parametric form by $r(t),\varphi(t),z(t)$.
That is 
\begin{eqnarray}
\dot{r}(t) & = & 0\nonumber \\
r(t)\dot{\varphi}(t) & = & \frac{\beta a^{2}}{r^{3}}\\
\dot{z}(t) & = & \frac{a^{2}\sqrt{r^{2}+\beta^{2}}}{r^{3}}\nonumber 
\end{eqnarray}
After straightforward calculations, the solutions of (\ref{geod})
are obtained as: 
\begin{eqnarray}
r(t) & = & r_{o}\nonumber \\
\varphi(t) & = & \frac{\beta a^{2}}{r_{o}^{4}}t+\varphi_{o}\label{parametric}\\
z(t) & = & \frac{a^{2}\sqrt{r_{o}^{2}+\beta^{2}}}{r_{o}^{3}}t+z_{o},\nonumber 
\end{eqnarray}
where $r_{o}$, $\varphi_{o}$ and $z_{o}$ are integration constants.
It is clear that the set of equations (\ref{parametric}) describes
a helix of radius $r_{o}$ and pitch $2\pi\frac{r_{o}\sqrt{r_{o}^{2}+\beta^{2}}}{\beta}$.
Notice that, in the absence of the defect, $r=r_{o}$, $\varphi=\varphi_{o}$
and $z=z_{o}+const\cdot t$, which represents a straight line along
the $z$-axis.

Now, we turn our attention to vector potential $\vec{A}$ and the
scalar potential $V$. These latter can be obtained from the electric
and magnetic fields given by (\ref{ansatz}) assorted by an appropriate
gauge condition. For convenience, the Coulomb gauge is used in all
that follows, so that potentials are going to be obtained from: 
\begin{eqnarray}
\vec{B} & = & \overrightarrow{\nabla}\wedge\vec{A}\label{pot1}\\
\vec{E} & = & -\frac{\partial\vec{A}}{\partial t}-\overrightarrow{\nabla}V\label{pot2}\\
\overrightarrow{\nabla}.\vec{A} & = & 0\label{pot3}\\
\triangle V & = & 0\label{pot4}
\end{eqnarray}

Using the amplitudes of the fields as prescribed by (\ref{Efield})
and (\ref{Bfield}), the previous set of equations gives for (\ref{pot1})
\begin{eqnarray}
\frac{1}{r}\frac{\partial A_{z}}{\partial\phi}-\frac{\partial A_{\phi}}{\partial z} & = & \frac{a}{r}e^{i(m\phi+kz-\omega t)}\label{pot1r}\\
\frac{\partial A_{r}}{\partial z}-\frac{\partial A_{z}}{\partial r} & = & \frac{a}{r^{2}}\sqrt{r^{2}+\beta^{2}}e^{i(m\phi+kz-\omega t)}\label{pot1phi}\\
\frac{\partial}{\partial r}(rA_{\phi})-\frac{\partial A_{r}}{\partial\phi} & = & -\frac{\beta a}{r}e^{i(m\phi+kz-\omega t)}\label{pot1z}
\end{eqnarray}
This system suggests that each of the unknown functions $A_{r}$,$A_{\phi}$
and $A_{z}$ are linear with respect to the factor $e^{i(m\phi+kz-\omega t)}$.
In particular, this implies that: 
\begin{equation}
\frac{\partial\vec{A}}{\partial t}=-i\omega\vec{A}
\end{equation}
Therefore, (\ref{pot2}) gives the system: 
\begin{eqnarray}
i\omega A_{r}-\frac{\partial V}{\partial r} & = & \frac{a}{r}e^{i(m\phi+kz-\omega t)}\label{pot2r}\\
i\omega A_{\phi}-\frac{1}{r}\frac{\partial V}{\partial\phi} & = & -\frac{a}{r^{2}}\sqrt{r^{2}+\beta^{2}}e^{i(m\phi+kz-\omega t)}\label{pot2phi}\\
i\omega A_{z}+\frac{\partial V}{\partial z} & = & \frac{\beta a}{r^{2}}e^{i(m\phi+kz-\omega t)}\label{pot2z}
\end{eqnarray}
This in turn implies that the scalar potential $V$ is linear with
respect to the factor $e^{i(m\phi+kz-\omega t)}$. Bearing in mind
there is a similar property for the vector potential, this strongly
suggests that for both potentials, variables can be separated in the
following way: 
\begin{eqnarray}
A_{r} & = & a_{r}(r)e^{i(m\phi+kz-\omega t)}\label{hypar}\\
A_{\phi} & = & a_{\phi}(r)e^{i(m\phi+kz-\omega t)}\label{hypap}\\
A_{z} & = & a_{z}(r)e^{i(m\phi+kz-\omega t)}\label{hypaz}\\
V & = & v(r)e^{i(m\phi+kz-\omega t)}\label{hypv}
\end{eqnarray}
Therefore, expressing the Coulomb gauge equations (\ref{pot3})-(\ref{pot4}),
it comes straightforwardly that: 
\begin{eqnarray}
\frac{1}{r}\frac{\partial}{\partial r}\left(rA_{r}\right)+\frac{1}{r}\frac{\partial A_{\phi}}{\partial\phi}+\frac{\partial A_{z}}{\partial z}=0\label{pot-vec-1}\\
\frac{1}{r}\frac{\partial}{\partial r}\left(r\frac{\partial V}{\partial r}\right)+\frac{1}{r^{2}}\frac{\partial^{2}V}{\partial\phi^{2}}+\frac{\partial^{2}V}{\partial z^{2}}=0\label{pot-vec-1}\\
\end{eqnarray}
Using (\ref{hypv}) in (\ref{pot-vec-1}), it comes that after some
algebra that: 
\begin{equation}
\frac{d^{2}v}{dr^{2}}+\frac{1}{r}\frac{dv}{dr}-\left(\frac{m^{2}}{r^{2}}+k^{2}\right)v=0\label{pot-vec-2}
\end{equation}
Performing the change in variable $X\leftrightarrow k\: r$, we can
rearrange the previous expression to get: 
\begin{equation}
X^{2}\frac{d^{2}v}{dX^{2}}+X\frac{dv}{dX}-\left(m^{2}+X^{2}\right)v=0\label{MBessel}
\end{equation}
The solutions of this equation are the modified Bessel functions $I_{\pm m}(X)$
and $K_{m}(X)$. As the electric field involves the divergence of
the scalar potential, it is natural to retain only the $K_{m}(X)$
functions so that the electromagnetic field vanishes at infinity.
Therefore, pluging $v(r)=K_{m}(k\: r)$ in eqs (\ref{pot2r})-(\ref{pot2z})
and using the ansatz (\ref{hypar})-(\ref{hypaz}), we are led to:
\begin{eqnarray}
a_{r}(r) & = & \frac{ik}{2\omega}\left(K_{m+1}(kr)+K_{m-1}(kr)+\frac{2a}{kr}\right)\label{exppotr}\\
a_{\phi}(r) & = & \frac{i}{\omega}\left(\frac{a}{r^{2}}\sqrt{r^{2}+\beta^{2}}-im\frac{K_{m}(kr)}{r}\right)\label{exppotphi}\\
a_{z}(r) & = & \frac{i}{\omega}\left(-\frac{\beta a}{r^{2}}+ikK_{m}(kr)\right)\label{exppotz}\\
\end{eqnarray}

The orbital angular momentum is defined from the vector potential
and the electric field by \cite{Jackson}: 
\begin{equation}
\vec{L}=\sum_{j=1}^{3}\frac{1}{\mu_{0}c^{2}}\int d^{3}xE_{j}\left(\vec{x}\wedge\vec{\nabla}\right)A_{j}
\end{equation}

The volume density of angular momentum, which at the point $\vec{R}$,
is given by 
\begin{equation}
\vec{M}=\vec{R}\times\vec{S}.
\end{equation}
A straightforward calculation gives 
\begin{equation}
\vec{M}=-\frac{z\beta a^{2}}{r^{3}}\vec{e}_{\hat{r}}-\frac{a^{2}\sqrt{r^{2}+\beta^{2}}}{r^{2}}\vec{e}_{\hat{\varphi}}+\frac{\beta a^{2}}{r^{2}}\vec{e}_{\hat{z}}.
\end{equation}

It is well-known that with a unit system in which $c=1$, the linear
momentum density identifies with the Poynting vector. Thus, the ratio
between the flux of angular momentum to that of energy across the
surface $z$=$const$ is given by: 
\[
L/P=\frac{\int_{\delta}^{\infty}dr\int_{\delta}^{2\pi}rd\varphi M_{z}}{\int_{\delta}^{\infty}dr\int_{\delta}^{2\pi}rd\varphi S_{z}}=\beta\frac{\int_{\delta}^{\infty}\frac{dx}{x}}{\int_{\delta}^{\infty}\frac{\sqrt{1+x^{2}}}{x^{2}}dx}
\]
where $\delta$ is an ultraviolet cut-off corresponding to a core
structure. In smectic liquid crystals, $\delta$ is of the order of
the average thickness of a layer \cite{DeGennes}, whereas in a cosmological
context, $\delta$ is about the inverse of the energy scale at which
the symmetry-breaking phase transition occurs \cite{BAllen}. Then
after simple manipulations, it comes that 
\begin{equation}
L/P=\beta=m/\omega
\end{equation}
The solution (\ref{ansatz}) has therefore a well-defined orbital
angular momentum. It has to be emphasized that even if this result
is derived from a simple solution of Maxwell's equations, it corresponds
to what is obtained for realistic laser modes (Laguerre-Gaussian beams
\cite{allen}).

Before closing this section a few remarks on the conservation laws
are in order. It is interesting to notice that the Poynting vector
(\ref{poynting}) obeys the conservation law $\vec{\nabla}\cdot\vec{S}=0$.
Furthermore, the transversal part of the Poynting vector is also divergence-free.
So, the beam intensity distribution does not change in the plane perpendicular
to the direction of propagation. In other words, it is a non-diffracting
beam (\cite{horak}). Also, the radial and azimuthal components of
the angular momentum density are symmetric about the $z$-axis. This
implies that integration over the beam profile leaves only the $\vec{e}_{\hat{z}}$
component. This is easily seen by writing $\vec{M}$ in terms of its
Cartesian components while keeping the cylindrical coordinates.

\section{Conclusion}

In this work, we investigated some features of electrodynamics in
the neighborhood of a screw dislocation. From the geometric treatment
of topological defects, it appears that the torsion induced by the
dislocation couples to the electromagnetic field in two ways. First,
it is responsible for a quantization of the modes, for which the allowed
frequencies depend only on the value of Burgers vector. This may be
of prime interest for several potential applications such as defect
sounding or X-ray filters or even the design on the heat rectifier
devices \cite{Maldovan}, due to the periodicity of the screw dislocation.
Second, the torsion forces the Poynting vector to spiral along the
direction of propagation, possibly endowing the electromagnetic field
with an orbital angular momentum. Such property is relevant in observational
cosmology as a signature of cosmic strings, but it also provides an
alternate approach to design optical tweezers from a simple (and tunable)
waveguide effect.

One of the main interests of this work is that it can be generalized
to other kinds of defects. Indeed, the differential forms formalism
provides the general process of dealing with electromagnetism in non-trivial
background geometries. Other kinds of line defects (edge dislocations,
disclinations) and even distributions of defects can be treated this
way, and one may expect strong couplings between the quantized modes
in this last case. This will be the object of a next paper.
\begin{acknowledgments}
F.M. is grateful to LEMTA for financial support during his stay there
and CNPq, CAPES (Brazilian agencies) for financial support. E.P. is
grateful to FAPEAL and CNPq (Brazilian agencies) for financial support.
The authors thank Pablo Vaveliuk and Dragi Karevski for fruitful discussions
related to part III.
\end{acknowledgments}
\global\long\def\theequation{A-\arabic{equation}}
 \setcounter{equation}{0} 

\section*{APPENDIX: Hodge duality}

In cylindrical coordinates, the 1-form basis writes: 
\begin{equation}
\mathcal{B}_{1}=\{\mbox{\boldmath\ensuremath{d}}t,~\mbox{\boldmath\ensuremath{d}}r,~\mbox{\boldmath\ensuremath{d}}\varphi,~\mbox{\boldmath\ensuremath{d}}z\}\label{1formbasis}
\end{equation}
In electrodynamics, the components of the Faraday 2-form that accounts
for the field write as \cite{warnick}:

\begin{equation}
\mbox{\boldmath\ensuremath{F}}=\frac{1}{2}F_{\mu\nu}\mbox{\boldmath\ensuremath{d}}x^{\mu}\wedge\mbox{\boldmath\ensuremath{d}}x^{\nu},
\end{equation}

\noindent where $\mbox{\boldmath\ensuremath{d}}x^{\mu}\wedge\mbox{\boldmath\ensuremath{d}}x^{\nu}$
are elements of the 2-form basis.

\begin{equation}
\mathcal{B}_{2}=\{\mbox{\boldmath\ensuremath{d}}\varphi\wedge\mbox{\boldmath\ensuremath{d}}z,~\mbox{\boldmath\ensuremath{d}}z\wedge\mbox{\boldmath\ensuremath{d}}r,~\mbox{\boldmath\ensuremath{d}}r\wedge\mbox{\boldmath\ensuremath{d}}\varphi,~\mbox{\boldmath\ensuremath{d}}r\wedge\mbox{\boldmath\ensuremath{d}}t,~\mbox{\boldmath\ensuremath{d}}\varphi\wedge\mbox{\boldmath\ensuremath{d}}t,~\mbox{\boldmath\ensuremath{d}}z\wedge\mbox{\boldmath\ensuremath{d}}t\}.\label{base2-formas}
\end{equation}

\noindent To translate the usual Maxwell's equations in terms of differential
forms, it is convenient to introduce the Hodge star operator $\star$.
This latter acts on a $p$-form in $n$-dimensional space and turns
it into the ($n-p$)-form that somehow completes the volume $n$-form.
Given the product of two p-forms $\rho$ and $\psi$ defined on an
oriented \textit{n}-manifold described by metric $g_{\mu\nu}$, then
the Hodge star operator is defined as: 
\begin{equation}
\rho\wedge\star\psi=\sqrt{\left|det(g_{\mu\nu})\right|}\left\langle \rho,\psi\right\rangle \mbox{\boldmath\ensuremath{d}}x^{1}\wedge..\mbox{\boldmath\ensuremath{d}}x^{n-p}\label{Hodge}
\end{equation}
Taking into account metric (\ref{metric}), the action of $\star$
on the 2-forms of $\mathcal{B}_{2}$ is then:

\begin{eqnarray}
\star(\mbox{\boldmath\ensuremath{d}}\varphi\wedge\mbox{\boldmath\ensuremath{d}}z) & = & -\frac{1}{r}\mbox{\boldmath\ensuremath{d}}r\wedge\mbox{\boldmath\ensuremath{d}}t,\\[0.2cm]
\star(\mbox{\boldmath\ensuremath{d}}z\wedge\mbox{\boldmath\ensuremath{d}}r) & = & -\frac{r^{2}+\beta^{2}}{r}\mbox{\boldmath\ensuremath{d}}\varphi\wedge\mbox{\boldmath\ensuremath{d}}t-\frac{\beta}{r}\mbox{\boldmath\ensuremath{d}}z\wedge\mbox{\boldmath\ensuremath{d}}t,\\[0.2cm]
\star(\mbox{\boldmath\ensuremath{d}}r\wedge\mbox{\boldmath\ensuremath{d}}\varphi) & = & -\frac{\beta}{r}\mbox{\boldmath\ensuremath{d}}\varphi\wedge\mbox{\boldmath\ensuremath{d}}t-\frac{1}{r}\mbox{\boldmath\ensuremath{d}}z\wedge\mbox{\boldmath\ensuremath{d}}t,\\[0.2cm]
\star(\mbox{\boldmath\ensuremath{d}}r\wedge\mbox{\boldmath\ensuremath{d}}t) & = & r\mbox{\boldmath\ensuremath{d}}\varphi\wedge\mbox{\boldmath\ensuremath{d}}z,\\[0.2cm]
\star(\mbox{\boldmath\ensuremath{d}}\varphi\wedge\mbox{\boldmath\ensuremath{d}}t) & = & \frac{1}{r}\mbox{\boldmath\ensuremath{d}}z\wedge\mbox{\boldmath\ensuremath{d}}r-\frac{\beta}{r}\mbox{\boldmath\ensuremath{d}}r\wedge\mbox{\boldmath\ensuremath{d}}\varphi,\\[0.2cm]
\star(\mbox{\boldmath\ensuremath{d}}z\wedge\mbox{\boldmath\ensuremath{d}}t) & = & \frac{r^{2}+\beta^{2}}{r}\mbox{\boldmath\ensuremath{d}}r\wedge\mbox{\boldmath\ensuremath{d}}\varphi-\frac{\beta}{r}\mbox{\boldmath\ensuremath{d}}r\wedge\mbox{\boldmath\ensuremath{d}}z.
\end{eqnarray}

\noindent On the elements of the 3-form basis $\mathcal{B}_{3}$,
the action of Hodge's star operator is

\begin{eqnarray}
\star(\mbox{\boldmath\ensuremath{d}}r\wedge\mbox{\boldmath\ensuremath{d}}\varphi\wedge\mbox{\boldmath\ensuremath{d}}t) & = & -\frac{1}{r}\mbox{\boldmath\ensuremath{d}}z-\frac{\beta}{r}\mbox{\boldmath\ensuremath{d}}\varphi,\\[0.2cm]
\star(\mbox{\boldmath\ensuremath{d}}z\wedge\mbox{\boldmath\ensuremath{d}}r\wedge\mbox{\boldmath\ensuremath{d}}t) & = & -\frac{r^{2}+\beta^{2}}{r}\mbox{\boldmath\ensuremath{d}}\varphi-\frac{\beta}{r}\mbox{\boldmath\ensuremath{d}}z,\\[0.2cm]
\star(\mbox{\boldmath\ensuremath{d}}\varphi\wedge\mbox{\boldmath\ensuremath{d}}z\wedge\mbox{\boldmath\ensuremath{d}}t) & = & -\frac{1}{r}\mbox{\boldmath\ensuremath{d}}r,\\[0.2cm]
\star(\mbox{\boldmath\ensuremath{d}}r\wedge\mbox{\boldmath\ensuremath{d}}\varphi\wedge\mbox{\boldmath\ensuremath{d}}z) & = & -\frac{1}{r}\mbox{\boldmath\ensuremath{d}}t.
\end{eqnarray}



\begin{thebibliography}{10}
\bibitem{foot} The explanation for that comes from the fact that
we have $\vec{v}=v^{\mu}\vec{e}_{\mu}=v^{\hat{\mu}}\vec{e}_{\hat{\mu}}$.
\end{thebibliography}
\end{document}